\documentclass[]{spie}  

\usepackage{lineno}

\usepackage{amsmath,amsfonts,amssymb}
\usepackage{graphicx}
\usepackage[colorlinks=true, allcolors=blue]{hyperref}
\usepackage{subfiles}
\usepackage{placeins}

\title{In-lab and On-sky Closed-loop Results of Adaptive Secondary Mirrors with TNO's Hybrid Variable Reluctance Actuators}

\author[a]{Ruihan Zhang}
\author[b]{Max Baeten}
\author[a]{Mark R. Chun}
\author[a]{Ellen Lee}
\author[a]{Michael Connelley}
\author[c]{Olivier Lai}
\author[b]{Stefan Kuiper}
\author[a]{Alan Ryan}
\author[b]{Arjo Bos}
\author[d]{Rachel Bowens-Rubin}
\author[d]{Philip M. Hinz}
\affil[a]{Institute for Astronomy, University of Hawai'i at Manoa, 640 N Aohoku Pl, Hilo, USA}
\affil[b]{TNO, Stieltjesweg 1, NL-2628 CK Delft, Netherlands}
\affil[c]{Observatoire de la Cote d’Azur, 96 Bd de l'Observatoire, 06300 Nice, France}
\affil[d]{UC Santa Cruz, 1156 High St, Santa Cruz CA, USA}

\authorinfo{Further author information: (Send correspondence to R.Z.)\\R.Z.: E-mail: rzhang9@hawaii.edu, Telephone: +1 309 531 3014\\  M.C.: E-mail: markchun@hawaii.edu, Telephone: +1 808 932 2317}

\begin{document} 
\maketitle

\begin{abstract}
We performed closed-loop lab testing of large-format deformable mirrors (DMs) with hybrid variable reluctance actuators. TNO has been developing the hybrid variable reluctance actuators in support for a new generation of adaptive secondary mirrors (ASMs), which aim to be more robust and reliable. Compared to the voice coil actuators, this new actuator technology has a higher current to force efficiency, and thus can support DMs with thicker facesheets. Before putting this new technology on-sky, it is necessary to understand how to control it and how it behaves in closed-loop. We performed closed-loop tests with the Shack-Hartmann wavefront sensor with three large-format deformable mirrors that use the TNO actuators: DM3, FLASH, and IRTF-ASM-1 ASM. The wavefront sensor and the real-time control systems were developed for the NASA Infrared Telescope Facility (IRTF) and the UH 2.2-meter telescope ASMs. We tested IRTF-ASM-1 on-sky and proved that it meets all of our performance requirements. This work presents our lab setup for the experiments, the techniques we have employed to drive these new ASMs, the results of our closed-loop lab tests for FLASH and IRTF-ASM-1, and the on-sky closed-loop results of IRTF-ASM-1 ASM. 
\end{abstract}


\section{INTRODUCTION}
\label{sec:intro}  
Adaptive optics (AO) technology has benefited the field of astronomy for the past few decades providing higher image resolutions compared to traditional telescopes. \cite{Beckers1993} Pockets of air in the atmosphere have different temperatures and thus varying indices of refraction. Light from astronomical objects entering the earth's atmosphere gets distorted by the atmosphere as wind blows and convection takes place. An AO system uses its wavefront sensor (WFS) to measure the distortion caused by the atmosphere, calculates a set of corresponding commands with its real-time controller (RTC), and applies the commands to a deformable mirror (DM) that physically bends to negate the distortions caused by the atmosphere. This makes the incoming wavefront more coherent, creating sharper images at the focal plane. To keep up with the atmosphere, AO systems typically run with loop rates on the order of 1000 Hz.

An adaptive secondary mirror (ASM) can be used as the DM in an AO system. ASMs are situated inside the telescope, replacing the conventional static secondary mirror. The major advantages to using an ASM compared to a traditional AO system that wholly sits behind the telescope are: when conjugated to the ground layer, ASMs are capable of improving image quality at all wavelengths over a large field of view based on studies done by Gemini\cite{Szeto2006} and Subaru\cite{Subaru2020}. Especially at Maunakea, the ground layer is found to be confined and is the major contributor to optical turbulence \cite{Roddier1990,Marks1996,Chun2009}. In addition, the ASM is part of the telescope and is seen by all instruments on the telescope, all the instruments on the telescope may be benefited by the ASM. ASM-based AO systems have low thermal emissitivity and high optical throughput because extra relay optics are not needed to perform the AO correction. 

ASMs have proven successful at various observatories, such as the MMT \cite{Wildi2003},the Large Binocular Telescope \cite{Riccardi2010,Esposito2010}, the VLT \cite{Biasi2012}, and the Magellan Telescope\cite{Close2018}. These ASMs have realized the promises of providing correction over a large field of view with low thermal emissivity and high throughput. They all used the same actuator technology -- the voice coil actuators. These ASMs are complex, costly, and they have been difficult to operate and maintain \cite{Christou2014}. The choice of voice coil actuators drives many characteristics of these ASMs. With a relatively small force output ($\sim$1N) output, a think and delicate facesheet is required which is difficult to manufacture and handle. The thin facesheets drive the ASMs to high actuator densities so the facesheets do not sag between actuators. The weak force output also means the actuators are working at non-linear regimes, so capacitive sensors are required for every single actuator to know the actual shape of the ASM. These ASMs use a lot of energy and generate a large amount of heat that requires a complex cooling system. A new generation of ASMs is being developed using the hybrid variable reluctance (HVR) actuators invented by TNO \cite{Chun2022,Jonker2022,BRubin2021}. The HVR actuators have a force output that is about an order of magnitude larger than the voice coil actuators, allowing ASMs to have thicker facesheets and be operate in the linear regime. Operating in the linear regimes means the ASM can be used in open-loop mode with no sensor feedback. Furthermore, the facesheet is directly bonded to each actuator which improves robustness and ease of operation.

The first ASM to go on-sky using the HVR actuator technology is the IRTF-ASM-1 at the NASA Infrared Telescope Facility (IRTF). It is a 3-ringed 36 actuator (6+12+18) ASM with a diameter of 245 mm. Its convex facesheet is 3.3 mm thick and its radial actuator pitch is 37 mm \cite{Lee2024, Bos2024}. IRTF-ASM-1 uses the UH88 style actuators, which are also used in FLASH \cite{BRubin2021} (a flat DM with 19 actuators serving as a prototype for IRTF-ASM-1) and the UH88 ASM \cite{Chun2022}. Before IRTF-ASM-1 goes on-sky, it is imperative to characterize the ASM in the lab. During this process, we developed and assembled the DM electronics, the WFS, and the RTC to drive IRTF-ASM-1. We also experimented with multiple different controllers for driving the AO loop to achieve optimal dynamic performance for this style of ASMs. Due to the availability of the different ASMs we were working with, we first started the WFS and RTC integration in July 2023 with DM3.  DM3 is a prototype DM using the same HVR actuator as UH88 but with a different way of mounting. The actuators are mounted in rows where all actuators in one row are integrated in an aluminum beam. This makes only grid layouts possible. The UH88 style actuators are stand-alone units that can be mounted in any desired orientation. Initial testing with DM3 was useful because it uses the same electronics as FLASH and IRTF-ASM-1, and it is also a flat mirror with the same diameter as FLASH. Then, using the same setup, we performed closed-loop tests with FLASH once it arrived in Hawaii in December 2023. Finally, we closed the AO loop with IRTF-ASM-1 using the same RTC and WFS in the lab after its arrival in February 2024. IRTF-ASM-1 saw its on-sky first light in April 2024. In this work, we present the lab setup for FLASH and its Shack-Harmann WFS design (Section \ref{subsec:FLASHsetup}). We also show the testing instrument optical layout we used for IRTF-ASM-1 in-lab and on-sky (Section \ref{subsec:bbDesign}). In Section \ref{sec:labResults}, we showcase the closed-loop performance of FLASH and IRTF-ASM-1 using different loop controllers. Lastly, we present the on-sky dynamical performance of IRTF-ASM-1 in Section \ref{sec:onsky}.

\section{Lab SETUP}
\label{sec:labSetup}

\subsection{Wavefront sensor design and optical setup for FLASH}
\label{subsec:FLASHsetup}

\begin{figure}
    \centering
    \includegraphics[width=0.65\linewidth]{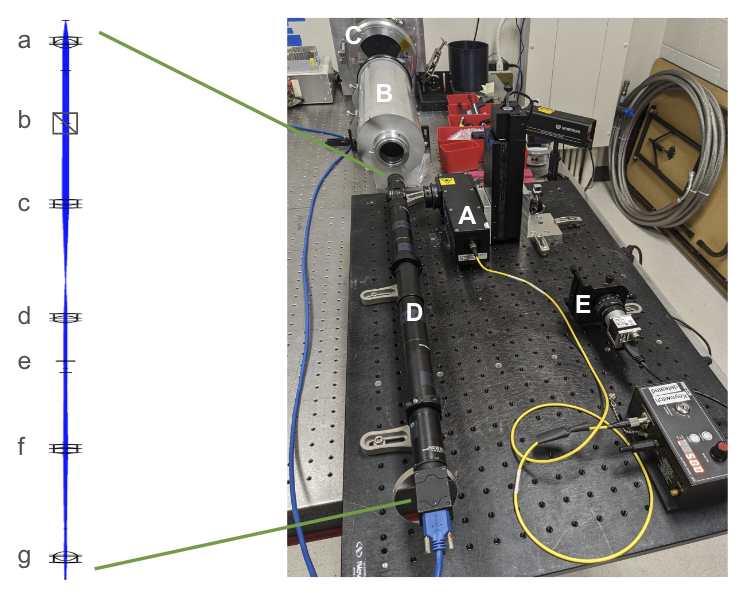}
    \caption{Optical setup for testing FLASH. The major components are: A) R-cube illuminator, B) 5-in lens/beam expander, C) FLASH, D) Single conjugate AO (SCAO) Shack-Hartmann WFS, E) HASO WFS. Inside the SCAO WFS$^D$, the optical components are: a. collimating lens, b. 50:50 beamsplitter, c.d. relay lenses, e. lenslet array, f.g. reimaging lenses. }
    \label{fig:lab_FLASH}
\end{figure}

The lab setup we used to test FLASH is shown in Figure \ref{fig:lab_FLASH}. The R-cube illuminator is a commercial product from Phasics that takes 625 nm light from a fiber source, collimates it and sends it out with a cube beamsplitter. The R-cube serves as the light source in all of the lab setups. The R-cube illuminates the DM via a beam expander. The beam is expanded to a 127 mm diameter beam by an in-hand f/5 lens. This illuminates most of FLASH (152 mm diameter).

After the light hits FLASH, it reflects back through the system and gets collimated again by the collimating lens at the entrance of the WFS tube. Then half of the remaining light transmits through the beamsplitter and gets resized by lenses (c) and (d) to illuminate a desired number of subapertures on the lenslet array. The lenslet array sits at a plane conjugated to the DM. The spot pattern formed by the lenslet array is then resized by lenses (f) and (g) to get the desired number of pixels per subaperture on the detector (FLIR Blackfly CMOS camera) that is at the end of the WFS tube. The SCAO Shack-Hartmann WFS has 12x12 subapertures and 10 pixels per subaperture. All the lenses in it are off the shelf achromatic doublets to mitigate chromatic aberration since we are taking the WFS on-sky. The optics for the WFS are in 1-inch lens tubes cut to specific lengths, which makes the assembly of the WFS very easy and precise. All parts in the WFS are on-shelf commercial components. This demonstrates an economic route to custom built WFSs that are suitable to a given camera/AO system that has few constraints, such as space.

Half of the light coming back from FLASH gets directed to the HASO WFS (a commercial product from Imagine Optic). The DM is conjugated to the HASO WFS. The HASO WFS is a high-order Shack-Hartmann system with a detector size of 5.3 mm x 7 mm using 50 x 68 subapertures. The purpose of the HASO WFS is to monitor the higher order aberrations that are not seen by the 12x12 WFS and not corrected with FLASH. We use the 12x12 WFS to drive the DM under test.

Other essential parts to the AO system are the driving electronics and the RTC. The DMs are driven by custom voltage to current amplifiers made by TNO. A commercial Linux machine is used as the RTC that communicates between the DM and the WFS using in-house RTC code.

\subsection{Design of the wavefront sensor package for IRTF-ASM-1}
\label{subsec:bbDesign}

\begin{figure}
    \centering
    \includegraphics[width=0.85\linewidth]{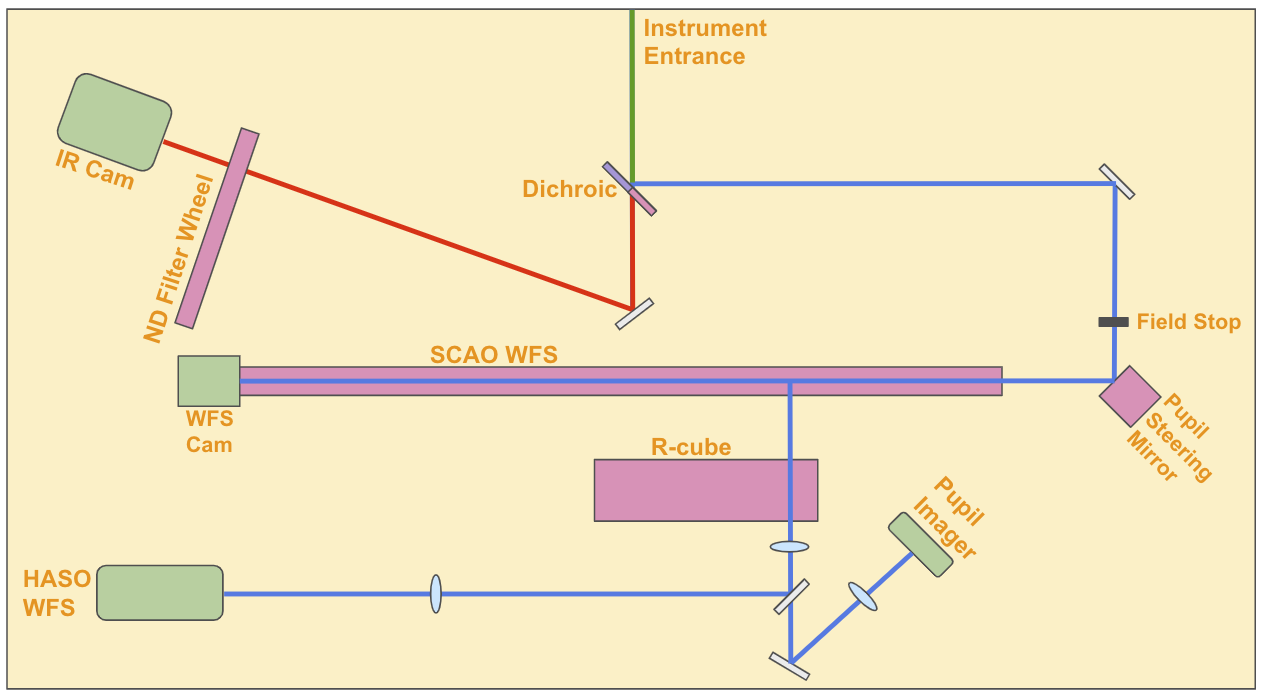}
    \caption{Scaled schematic diagram of the Test Instrument for IRTF-ASM-1.}
    \label{fig:bbIRTF1}
\end{figure}

There are more components to the wavefront sensor package for IRTF-ASM-1 (Figure \ref{fig:bbIRTF1}) because it is used for both in-lab and on-sky testing. In addition to the optical components described in Section \ref{subsec:FLASHsetup}, two major components added are: 1) a point spread function (PSF) imager and 2) a pupil steering mechanism.

The f/38 Cassegrain focus of IRTF is about 700 mm behind the telescope interface plate. We designed the Test Instrument with these conditions in mind. The dichroic behind the instrument entrance reflects visible light to the arm with the WFSs at the bottom of Figure \ref{fig:bbIRTF1} and transmits the IR wavelengths to the arm with the IR camera at the top left. The IR camera is a QHY990 camera sensitive from 400 nm to 1.7 microns. A 1.62 micron narrow bandpass is mounted in the camera to limit the light to longer wavelengths. The IR camera serves as a scoring camera that allows us to image the PSF and quantify the AO system performance. A neutral density (ND) filter wheel sits in front of the IR camera making sure the IR camera does not get saturated. The telescope focus falls on two locations in our instrument: one at the IR camera, the other at the field stop. 

A pupil steering mirror is located after the field stop in the wavefront sensor path to actively steer/maintain the pupil image on the wavefront sensor lenslet array. We added this mechanism to compensate for flexure between the telescope (in particular the secondary) and our instrument that would misregister the ASM and the WFS. The pupil steering mirror is used in closed-loop with a pupil imaging camera. 

When the wavefront sensor package is used in the lab, the R-cube is used as the illumination source. To go from in-lab to on-sky, we only need to change the collimating lens in the wavefront sensor tube since the returning beam in the lab is f/13, whereas light from the telescope would be f/38. Using the setup described above, we performed closed-loop experiments for IRTF-ASM-1.

\section{LAB PERFORMANCE AND CONTROL SERVO TUNING}
\label{sec:labResults}

\subsection{system identification}
\label{subsec:sysId}
In order to find the optimal closed loop controller for an ASM and its corresponding components (WFS \& RTC), the frequency response of the system is required. We tested the FLASH and the IRTF-ASM-1 system by sending white noise actuator commands in open-loop to a single actuator while recording slope values measured by the wavefront sensor. The slope measurements were converted into an output actuator response for that actuator using the interaction matrix of the system. By comparing the input noise signal and the measured system response, we obtain the single input single output frequency response of the system in both phase and amplitude for each actuator.

\begin{figure}[!htb]
    \centering
    \includegraphics[width=0.8\linewidth]{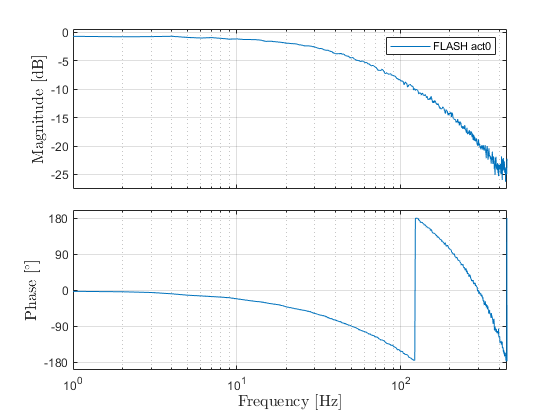}
    \caption{System frequency response for FLASH actuator 0.}
    \label{fig:FLASH_Bode}
\end{figure}

\begin{figure}[!htb]
    \centering
    \includegraphics[width=0.8\linewidth]{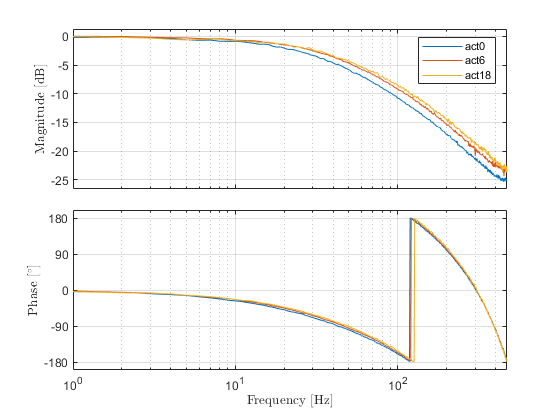}
    \caption{System frequency response for IRTF-ASM-1 actuator 0 (inner ring), 6 (middle ring) and 18 (inner ring)}
    \label{fig:IRTF_Bode}
\end{figure}

The identification measurements on actuator 0 of FLASH, as shown in Figure \ref{fig:FLASH_Bode} were measured with the system running at 895 Hz. The other actuators in FLASH all show similar responses and are therefore omitted.
A model has been fitted to this frequency response function.  We find it has a 2.5-sample delay that can be attributed to 0.5 step of sampling time, 1 step for image readout, and another step to perform calculations and apply actuator commands. Reducing this delay or improving the loop rate would improve the closed loop performance. The limiting factor of the loop rate is the readout time of the SCAO WFS camera. As an initial step, we sacrificed the dynamic range of the camera pixels going from 12-bit to 10-bit to obtain a higher loop speed of 998 Hz. Therefore the IRTF-ASM-1 system frequency response in Figure \ref{fig:IRTF_Bode} is measured at a higher loop rate of 998Hz.
The open loop frequency response for 3 out of 36 actuators is shown as they are all very similar. There is a variation at low frequencies at a level of about 15\%. Furthermore the phase drop for some actuators is higher than for others. Both of these variations are within predicted margins and are likely caused by fluctuations in actuator efficiency and in facesheet stiffness. The average low frequent gain per ring is 0.986, 1.00 and 1.036 respectively for the inner, middle and outer ring of actuators. The higher gain of the outer ring is the result of the lower facesheet at the outer edge. The faster phase drop of actuator 0 is the result of small variations in the actuators, if we select three different actuators than an actuator from the other two rings might show the fastest drop. All other actuators show phase drops within the same range. 

\FloatBarrier
\subsection{modal vs zonal control}
\label{subsec:ModVZon}

\begin{figure}[!htb]
    \centering
    \includegraphics[width=0.8\linewidth]{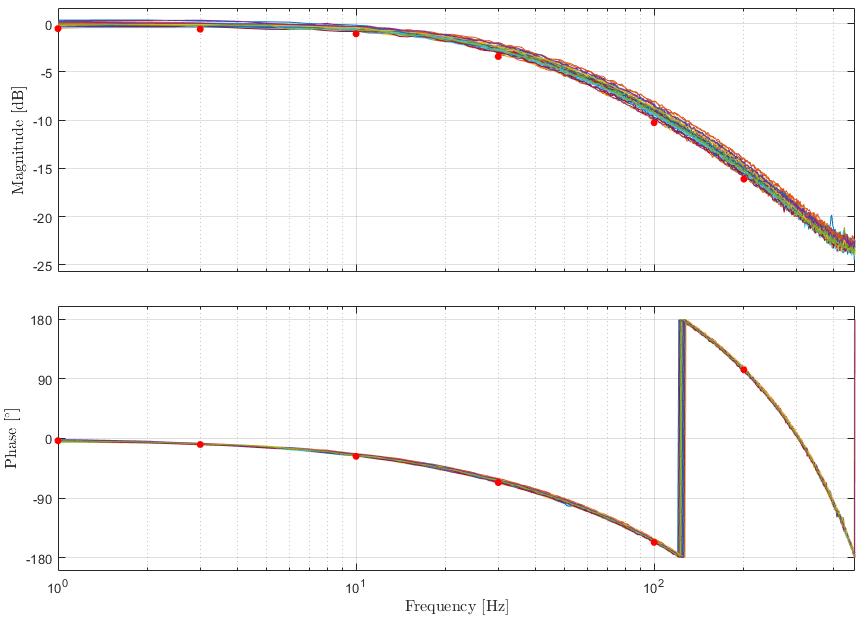}
    \caption{Bode diagram of IRTF-1 for Zernike mode Z3-Z37 with a verification with single sines (red dots) for Zernike mode Z5, primary oblique astigmatism}
    \label{fig:IRTF_ModalBode}
\end{figure}

A typical orthogonal basis set used by the community is the Zernike polynomials basis \cite{Noll1976}. Using influence functions of IRTF-1, we can convert the zonal commands to modal commands and vice versa. We repeated the system identification measurements by injecting noise sequentially on the first 36 Zernike modes, Z3 to Z37 (in Noll's ordering \cite{Noll1976}). As shown in Figure \ref{fig:IRTF_ModalBode}, they exhibit very similar dynamic behaviors since they share the same style of actuators, the same facesheet thickness, and were driven with the same WFS and RTC. Furthermore, there are almost no resonance frequencies that can be attributed to Zernike Modes as those resonances have higher frequencies than what can be measured with the current loop rate. At 400 Hz there is a very minor resonance, as measured on Z03, Tilt.
To verify the frequency responses measured by injecting noise, a few verification measurements are performed. For primary Z5 oblique astigmatism, 6 measurements are performed where a single frequency (1, 3, 10, 30, 100, and 200 Hz) sine wave is injected and its response is shown in red dots.

\FloatBarrier
\subsection{control tuning}
\label{subsec:ctrlTun}

The first consideration for controlling the ASM is which basis to use. Common control schemes are to close the loop for each actuator or for each mode. The closed loop control scheme tested on IRTF-1 consists of 36 single input single output feedback controllers for each actuator. Traditionally AO systems use a simple Proportional-Integral (PI) controller. In this project we designed controllers not limited to PID using classical control theory. The WFS slopes are multiplied by the interaction matrix to obtain 36 actuator values. We want to keep those values as close to zero to obtain a flat wavefront. The feedback controller takes as input the error signal defined as the setpoint minus the actuator value (with the setpoint set to zero). 

To tune the controllers of the 36 control loops a first approach is to tune one controller that is used for all 36 control loops. It is possible to fine tune the controller for each actuator for optimal performance but the performance gain is minimal because the variations between actuaotor responses are small. To tune the controller the frequency responses of all actuators are taken into account, although in this paper we present only 3 actuators for clarity. In an iterative process, the parameters of the controller are changed while the closed loop, the open loop, and the sensitivity transfer functions are plotted predicting the performance of the closed loop controller. The design goals are to achieve a high open loop cross over frequency and high closed loop -3db bandwidth while meeting the robustness margins. Two controllers are designed for which the parameters are given in Table \ref{tab:control}.

\begin{table}[htbp]
\caption{Zonal controller C1 and C6 parameters and performance metrics. Performance metrics are provided as a range for the 36 actuotors.}
\label{tab:control}
\begin{center}
\begin{tabular}{||c c c c c c ||} 
 \hline
                            &           & C1            & C4            & C6        &   \\[0.75ex] 
 \hline\hline
 \underline{Controller:}    &           &               &               &           &       \\ 
 gain                       &           & 0.37          & 0.3           & 0.74      & [-]   \\ 
 integrator                 & zero      & 110           & 110           & 90        & [Hz]  \\ 
 Lead filter                & pole      & 44            & 100           & 120       & [Hz]  \\ 
                            & zero      & 400           & 600           & 950       & [Hz]  \\ 
 2nd order Low pass filter  & pole      & 400           & 400           & NA        & [Hz]  \\ 
                            & damping   & 0.7           & 0.7           & NA        & [-]   \\ 
 \hline\hline
 \underline{Performance:}   &           &               &               &           &       \\ 
 open loop crossover        &           & 34 - 47       & 25 - 31       & 44 - 58   & [Hz]  \\ 
 closed loop bandwidth (-3dB)&          & 81 - 137      & 47 - 61       & 95 - 163  & [Hz]  \\ 
 modulus margin $<$ 6dB     &           & 4.4 - 6.8     & 3.5 - 4.7     & 4.8 - 7.0 & [dB]  \\ 
 30\textdegree $<$ phase margin $<$ 60\textdegree & 
                                        & 53 - 65       & 45 - 58       &  38 - 50  & [\textdegree]\\ 
 gain margin $>$6dB         &           & 5.62 - 8.9    &10.9 - 13.6    & 5.7 - 9.0 & [dB]  \\ 

 \hline
\end{tabular}
\end{center}
\end{table}

To ease the switching of designed closed loop controllers a generic approach is taken where a state space controller is implemented that consists of four matrices A,B,C and D. Therefore no implementation of lead filters or low pass filters are required as all controller components together are converted to a single state space representation. Such a representation is also discretized for the specific loop rate. 

\begin{figure}[!htb]
    \centering
    \includegraphics[width=1.0\linewidth]{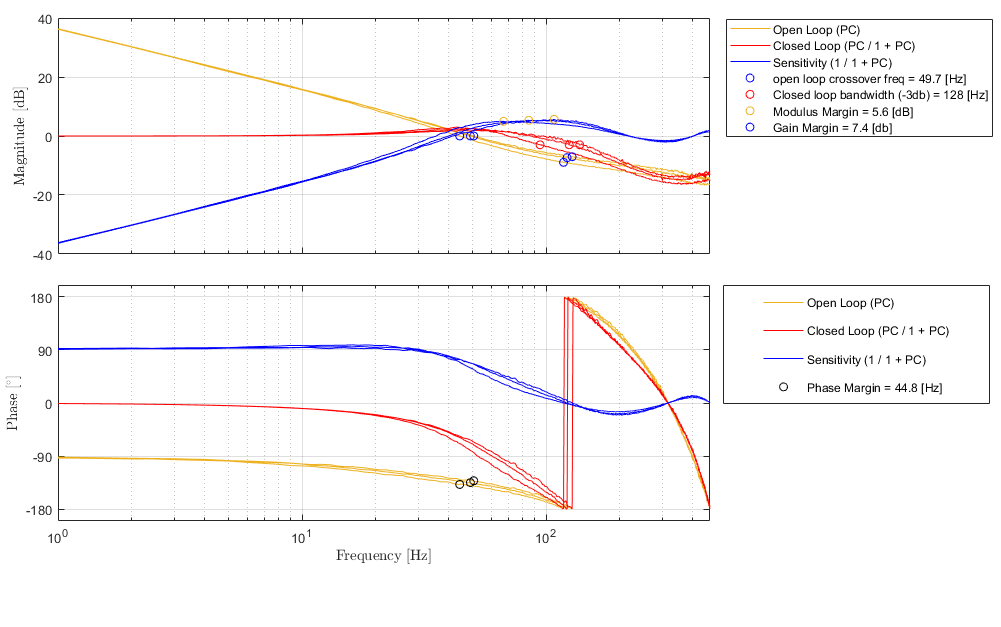}
    \caption{Control tuning bode plots, based on the measured system response P and the designed controller C. The bode plots are provided (with the same color) for three actuators 0, 6 and 18. The performance metrics from Table \ref{tab:control} are indicated with circles.}
    \label{fig:IRTF_contun}
\end{figure}
\FloatBarrier

\begin{figure}
    \centering
    \includegraphics[width=0.6\linewidth]{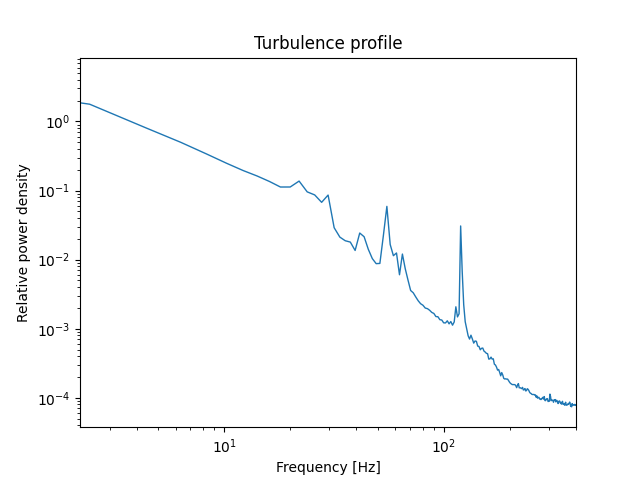}
    \caption{Example power spectrum of the atmospheric turbulence profile measured using the IRTF-1 AO system in open-loop at IRTF on Maunakea, April 26, 2024. The overall shape resembles typical atmospheric turbulence profiles. However, the sources that cause the spikes are not yet identified.}
    \label{fig:skyTurbulence}
\end{figure}

Error transfer functions should be examined with the input power spectrum taken into consideration since the end goal is to minimize residual error, which is the integral of the product of the input power spectrum and the error transfer function. The error that we are trying to correct with AO systems is caused by the atmosphere, which has a power spectrum that can be modeled using the Kolmogorov turbulence theory \cite{kolmogorov}. Most power in atmospheric turbulence is in the lower frequencies, and its strength falls with a -$\frac{5}{3}$ power-law as a function of frequency. An example of on-sky turbulence profile is shown in Figure \ref{fig:skyTurbulence}. Knowing that most power to be corrected is in the lower frequencies and there is not much to correct for in the higher frequencies, as long as the lower frequencies are well attenuated, there is margin for amplification of the higher frequencies as shown in the error transfer functions. Therefore, the error transfer function that provides optimal AO performance should have maximum attenuation at the lower frequencies and a high 0dB crossover frequency. There is always a part of the sensitivity that peaks above 0dB meaning that those frequencies are amplified. The maximum amplification should be as low as possible.

\FloatBarrier
\subsection{performance validation}
\label{subsec:perfVal}

\begin{figure}
    \centering
    \includegraphics[width=0.6\linewidth]{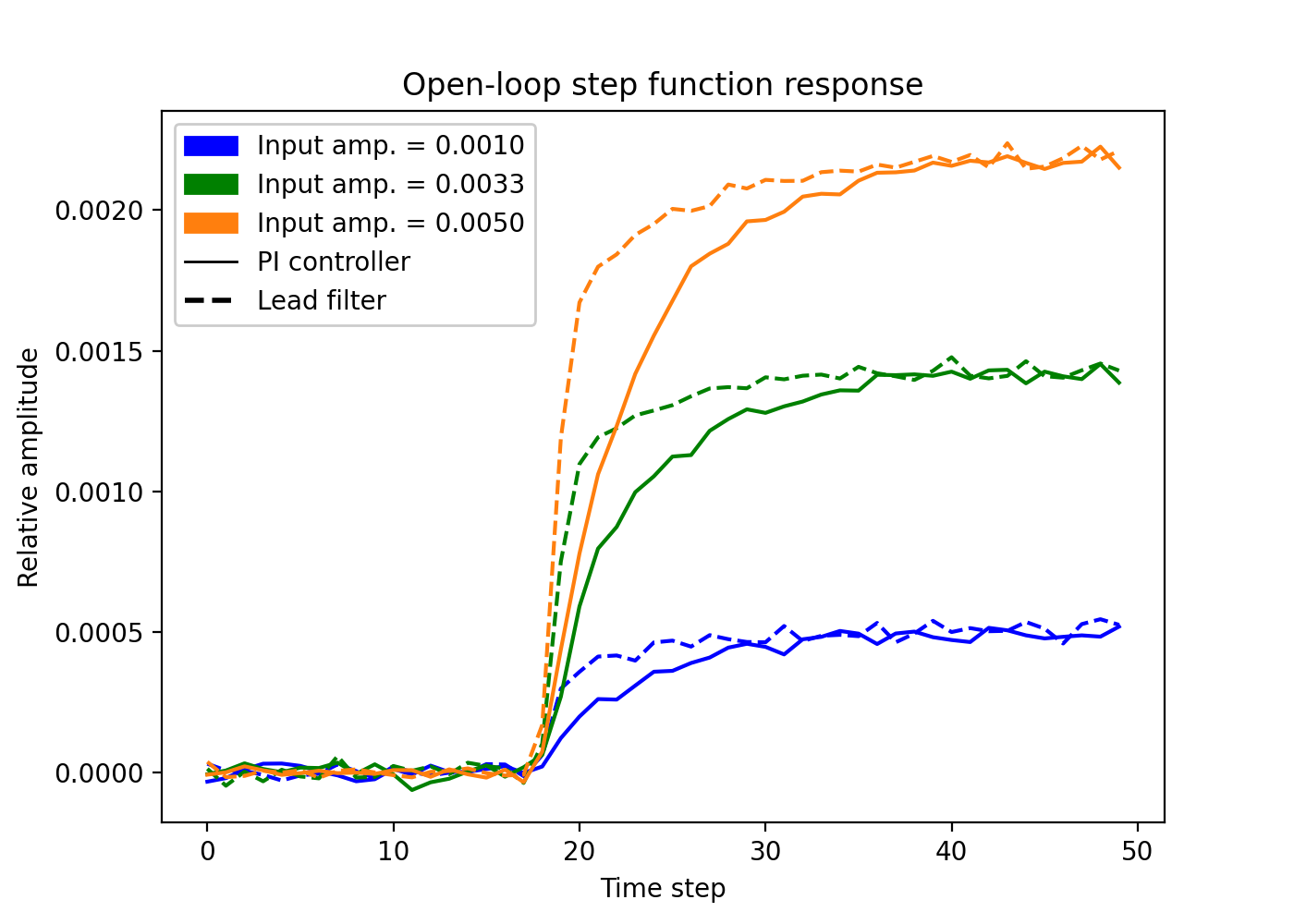}
    \caption{Open-loop step response of actuator 0 in IRTF-1. A lead filter significantly fastens the open-loop settling time of the actuator by about 37\%.}
    \label{fig:openStep}
\end{figure}

Another common metric widely used in the AO community that characterizes the performance of a DM is the settling time of its actuators. In other words, how fast does it take for an actuator to go from position A to position B. This can be shown in a step response curve of the actuator, which represents how the actuator behaves when it is commanded to move in the system. We examined the step response of actuator 0 in IRTF-1 in open-loop (Figure \ref{fig:openStep}). To study the open-loop response, we used the RTC to send direct actuator commands to actuator 0, then the WFS records the optical response caused by moving the actuator that deforms the surface of IRTF-1. 

Unlike some DMs that can move to position almost instantaneously and may be under-damped, such as the Micro-electro-mechanical (MEMS) DMs, the DMs made with the TNO UH88-style actuators need to be pushed into place, making it an over-damped system with a long open loop settling time. Closing the loop with a controller preferably with a lead filter included improves the settling time significantly. The settling time is defined here as the time used to reach and stay within a range of 20\% of the final value. In the first measurement, solid lines, the step is directly applied to the actuator. The second measurement, dashed lines, uses the same step but its fed through an open loop lead filter. Knowing these UH88 style actuators have a large inertia to start moving, the lead filter is designed to push the actuators harder initially to start the ASM moving faster and eases off the push in later steps to prevent excessive overshoot, thus achieving a shorter settling time. By using a lead filter, we shortened the settling time of actuator 0 by about 60\% with respect to a direct open-loop step. This was tested with three step sizes [0.0010, 0.0033, 0.0050] where the full dynamic range is -1 to 1 in arbitrary command space. These UH88-style actuators have a very large linear dynamic range of about 14 microns \cite{BRubin2021,Lee2024}. Only small incremental steps would be used when using the ASMs, thus the small magnitude of the test step sizes. The measured relative amplitude is half of the input amplitude because the output is normalized. For the direct step measurements, actuator 0 took [9, 10, 10] time steps to settle, whereas when a lead filter was used, only [4,4,4] time steps were needed. Since the system runs at 997 Hz, each time step is about 1 ms. In other words, in small amplitudes, the settling time of actuator 0 can be improved to about 4 ms by using lead filters. This explains why lead filters make an important component to the state space controllers for our systems, allowing them to outperform the PI controller. A few other actuators were also tested and they exhibit near identical behaviors to actuator 0.

\begin{figure}
    \centering
    \includegraphics[width=0.6\linewidth]{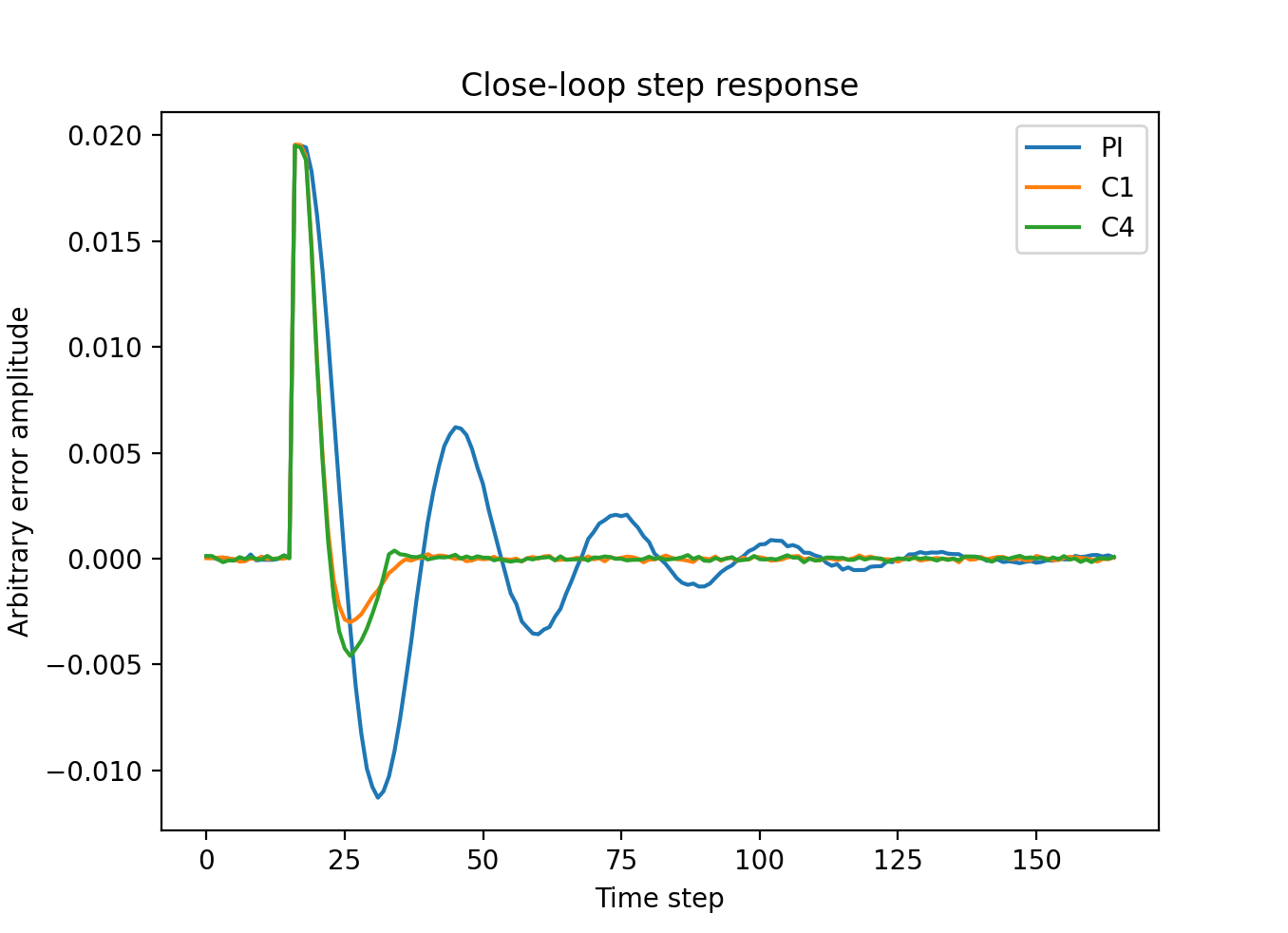}
    \caption{closed-loop step response of actuator 0 in IRTF-1. The two controllers, C1 and C4, both contain lead filters that improve the closed-loop settling time of the actuator by about 84\%.}
    \label{fig:closeStep}
\end{figure}

We also investigated the closed-loop response of the actuators, which also includes the response time of the RTC and is representing the response time of the whole AO system (Figure \ref{fig:closeStep}). We acquired the data by introducing slope offsets in the WFS, which suddenly introduces a large error in the closed-loop AO system. The system sees the error and moves the DM to compensate for the injected error. Using the PI controller, the system is under-damped, so the error signal oscillates for a long time before it settles near 0. On the other hand, when using state space controllers like C1 and C4 (a controller very similar to C6), the system is optimally damped. C1 and C4 both improve the closed-loop settling time by about 84\%. This also explains why the state space controllers show much better performance in error transfer functions.

\FloatBarrier

\begin{figure}
    \centering
    \includegraphics[width=0.6\linewidth]{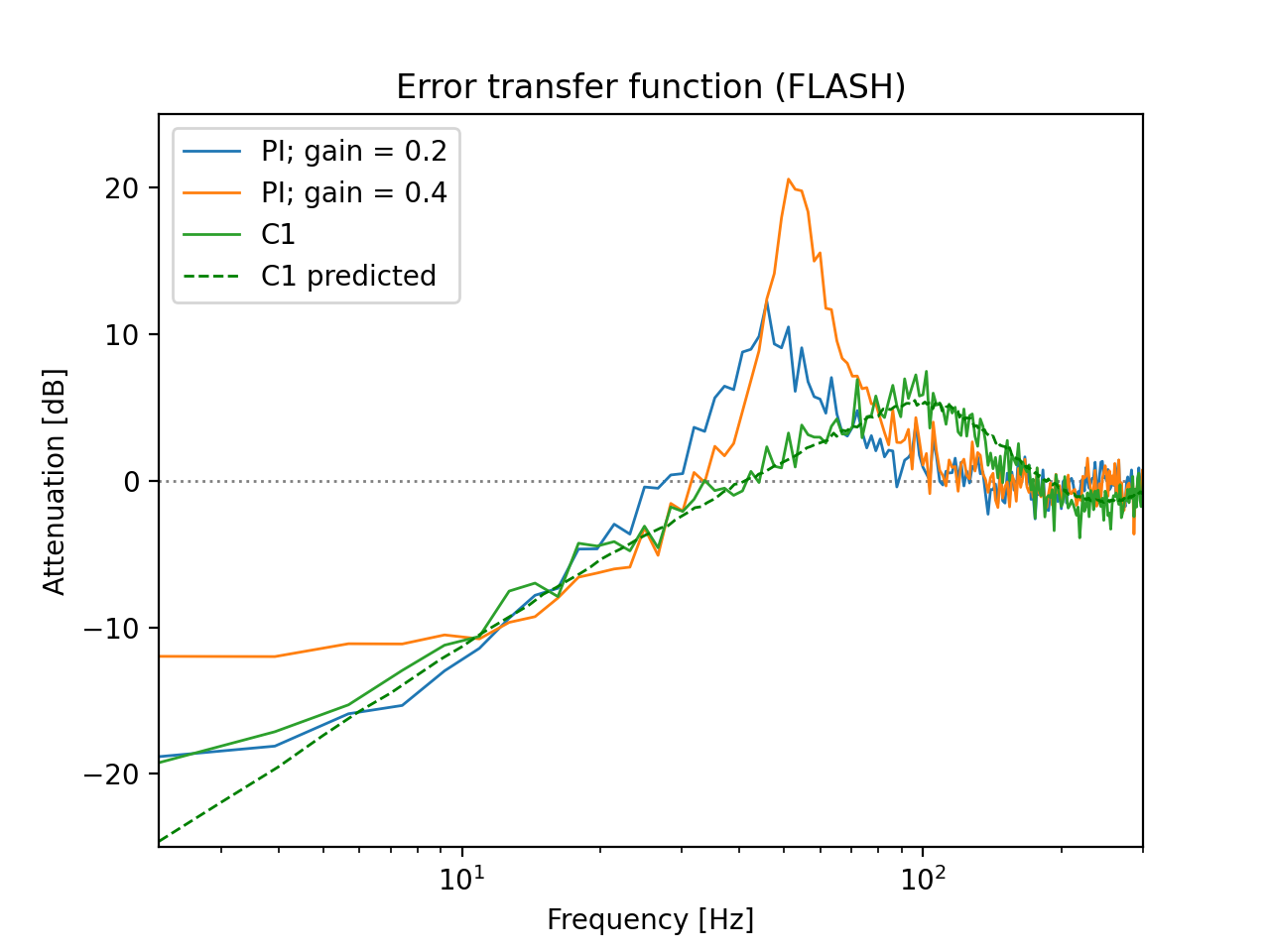}
    \caption{Error transfer functions of FLASH using different controllers. Frequencies below 0 dB (the gray dotted line) are being attenuated, and anything above it is amplified. The green dashed is the predicted performance of the system using controller C1, which matches well with the measured performance in the lab.}
    \label{fig:ETF_FLASH}
\end{figure}

We compared the performance of a typical PI controller and multiple state space controllers on the FLASH AO system. We took AO telemetry in closed-loop using the different controllers and compared them to the open-loop data, which is essentially white noise in camera signals since there is little turbulence in the lab. The ratio between the power spectra of closed-loop data and open-loop is the error transfer function or sensitivity function of the system. It showcases what frequencies are being attenuated by the closed-loop system (below the 0 dB line) and which are amplified. Looking at the error transfer functions of FLASH using the different controllers, controller C1, which is the best performing state space controller we tested, works much better than the conventional PI controller (Figure \ref{fig:ETF_FLASH}). 

\FloatBarrier
Here we conclude the dynamical testing of FLASH and IRTF-1 in the lab. The first observation with the IRTF-1 was conducted on April 24, 2024, marking the success of the first on-sky use of an ASM based on the TNO actuators.


\FloatBarrier
\section{ON-SKY PERFORMANCE OF IRTF-ASM-1 ON DYNAMICAL RESPONSE}
\label{sec:onsky}

\begin{figure}
    \centering
    \includegraphics[width=0.6\linewidth]{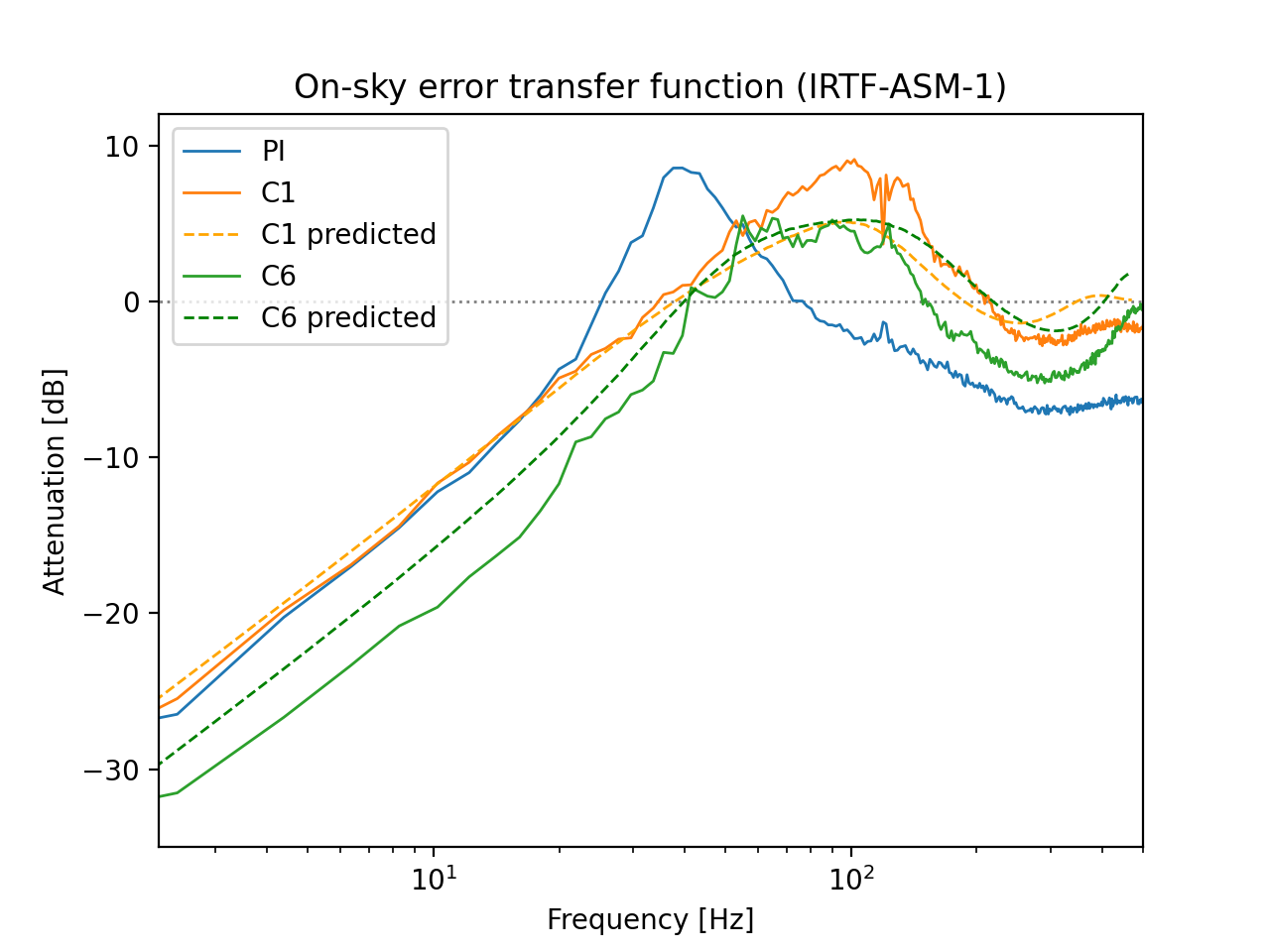}
    \caption{On-sky error transfer functions of IRTF-ASM-1 using different controllers. The data was taken on the nights of April 25/26, 2024.}
    \label{fig:ETF-onsky}
\end{figure}

On the nights of April 25 and 26, we got to test the effect that different controllers have on the performance of the AO system (Figure \ref{fig:ETF-onsky}). Similar to the lab results, the PI controller, which we ran using a proportional gain of 0.3, performed less well compared to the state space controllers C1 and C6. The 0 dB bandwidth measured on-sky for the PI controller, C1, and C6 are correspondingly [25, 35, 44] Hz. On the other hand, C6 seems to perform much better than C1 compared to their lab results. We examined the open-loop data used to generate the error transfer functions for C1 and C6, they were very similar and could not account for such large discrepancies. Therefore, the cause of such improvement needs to be further investigated. We plan to take more data with the different controllers under similar seeing conditions to study their effects since what we are currently have might be a single-time incident. One caveat to the on-sky error transfer functions is that unlike in the lab, the atmospheric turbulence is ever-changing. The AO telemetry we acquired include error signals that the wavefront sensor detects for 27000 steps running at 998 Hz, which corresponds to data of about 30 s. Each dataset was taken about 20 s after another due to the finite data transfer rate. Even though the open-loop telemetry and the closed-loop telemetry were taken one after the other with little time difference between the two, the open-loop data does not fully represent the atmospheric profile during the time that the closed-loop data was taken. This could cause inaccuracies in the on-sky error transfer functions and may be the reason that the error transfer functions are not converging to 0 dB at high frequencies. We could improve the accuracy of the on-sky error transfer functions by taking a set of closed-loop data sandwiched by two sets of open-loop telemetry and take the average of the two datasets as the open-loop spectrum.

\begin{figure}
    \centering
    \includegraphics[width=0.3\linewidth]{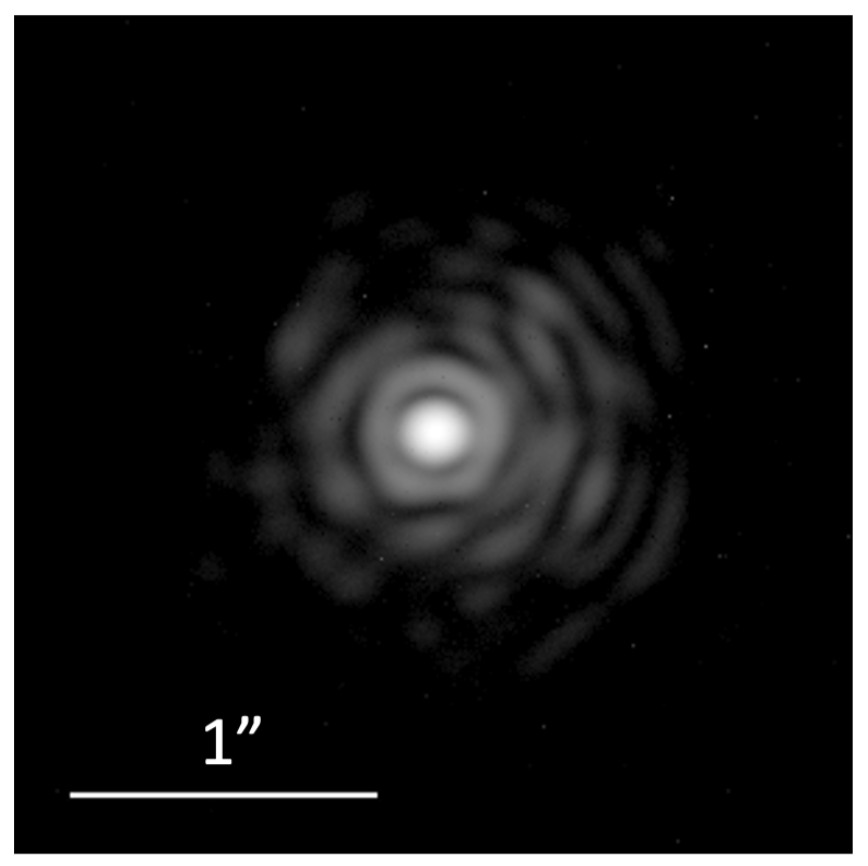}
    \caption{A long exposure (300 s) image of Arcturus (in an arcsinh scaling) at 1.62 microns with a Strehl ratio of 40\% running IRTF-ASM-1 in closed-loop (Lee et al. 2024)\cite{Lee2024}.}
    \label{fig:PSF}
\end{figure}

During the first observing run with IRTF-ASM-1, the best long-exposure we obtained was a diffraction limited image of Arcturus (Figure\ref{fig:PSF}). The PSF has a Strehl ratio of 40\% at the wavelength of 1.62 $\mu$m. It is a 300 s exposure made with 10 frames of 30 s exposures stacked together. The C6 state space controller was used while the image was taken. We look forward to further exploring the capabilities of ASMs with the TNO HVR actuators with future observing runs using IRTF-ASM-1 and the deployment of the University of Hawaii 2.2-meter ASM.

\section{DISCUSSION AND CONCLUSION}
\label{sec:conclusion}
We have successfully demonstrated the closed-loop AO performance of ASMs made with the novel HVR actuator technology both in-lab and on-sky. The IRTF-ASM-1 on NASA IRTF is capable of delivering diffraction limited images in H-band (1620 nm) with a Strehl raio of 35-40\% under sub-arcsecond seeing conditions. An important lesson we have learned through the closed-loop dynamical testing of these ASMs is that a slightly more sophisticated control servo than a PI loop (e.g. a leadfilter) improves the performance of the system.

Given the current frequency response of our AO system with IRTF-ASM-1, the RTC, and the as-built WFS, we were able to model the system as mentioned in Section \ref{subsec:sysId}. We have identified the WFS camera (off the shelf CMOS USB3 camera) as the limiting factor in the bandwidth of the system.   If we change to a more capable camera, we predict the 0dB bandwidth will improve to 50-85Hz (depending on a assumptions on the camera).   For the UH88ASM we plan to use the camera/wavefront sensor in the Robo-AO-2 system\cite{Baranec2024}. 

\acknowledgments 
 
This work is funded by an National Science Foundation (NSF-1910552) award. The authors would like to thank TNO and the NASA IRTF for the opportunity and assistance in making this project a success. Furthermore, we would like to thank The University California Santa Cruz for supplying a slumped facesheet and Huygens Optics for they spherical polishing of said mirror shell. Special thanks to LEMO for providing us connectors within a short timeframe, and BUMAX for helping us out solving a potential galling problem due to a manufacturing error. The on-sky experiments were performed at the Infrared Telescope Facility, which is operated by the University of Hawaii under contract 80HQTR19D0030 with the National Aeronautics and Space Administration. The authors want to express our gratitude towards the engineering team and the summit crew at IRTF for making the observing runs go smoothly. We also wish to recognize and acknowledge the very significant cultural role and reverence that the summit of Maunakea has always had within the Native Hawaiian community. We are most fortunate to have the opportunity to conduct observations from this mountain. 

\bibliography{report} 

\begin{thebibliography}{10}

\bibitem{Beckers1993}
{Beckers}, J.~M., ``{Adaptive Optics for Astronomy: Principles, Performance, and Applications},'' {\em Annual Review of Astron and Astrophysis}~{\bf 31},  13--62 (Jan. 1993).

\bibitem{Szeto2006}
{Szeto}, K., {Andersen}, D., {Crampton}, D., {Morris}, S., {Lloyd-Hart}, M., {Myers}, R., {Jensen}, J.~B., {Fletcher}, M., {Gardhouse}, W.~R., {Milton}, N.~M., {Pazder}, J., {Stoesz}, J., {Simons}, D., and {V{\'e}ran}, J.-P., ``{A proposed implementation of a ground layer adaptive optics system on the Gemini Telescope},'' in [{\em Society of Photo-Optical Instrumentation Engineers (SPIE) Conference Series}{\nolinebreak\hspace{0.1em}]},  {McLean}, I.~S. and {Iye}, M., eds., {\em Society of Photo-Optical Instrumentation Engineers (SPIE) Conference Series} {\bf 6269},  626958 (June 2006).

\bibitem{Subaru2020}
{Minowa}, Y., {Koyama}, Y., {Ono}, Y., {Tanaka}, I., {Hattori}, T., {Clergeon}, C., {Akiyama}, M., {Kodama}, T., {Motohara}, K., {Rigaut}, F., {d'Orgeville}, C., {Wang}, S.-Y., and {Yoshida}, M., ``{ULTIMATE-Subaru: enhancing the Subaru's wide-field capability with GLAO},'' in [{\em Advances in Optical Astronomical Instrumentation 2019}{\nolinebreak\hspace{0.1em}]},  {\em Society of Photo-Optical Instrumentation Engineers (SPIE) Conference Series} {\bf 11203},  112030G (Jan. 2020).

\bibitem{Roddier1990}
{Roddier}, F.~J., {Cowie}, L.~L., {Graves}, J.~E., {Songaila}, A., {McKenna}, D., {Vernin}, J., {Azouit}, M., {Caccia}, J.~L., {Limburg}, E.~J., {Roddier}, C.~A., {Salmon}, D.~A., {Beland}, S., {Cowley}, D.~J., and {Hill}, S., ``{Seeing at Mauna Kea: a joint UH-UN-NOAO-CFHT study},'' in [{\em Advanced Technology Optical Telescopes IV}{\nolinebreak\hspace{0.1em}]},  {Barr}, L.~D., ed., {\em Society of Photo-Optical Instrumentation Engineers (SPIE) Conference Series} {\bf 1236},  485--491 (July 1990).

\bibitem{Marks1996}
{Marks, R. D.}, {Vernin, J.}, {Azouit, M.}, {Briggs, J. W.}, {Burton, M. G.}, {Ashley, M. C.B.}, and {Manigault, J. F.}, ``Antarctic site testing - microthermal measurements of surface-layer seeing at the south pole,'' {\em Astron. Astrophys. Suppl. Ser.}~{\bf 118}(2),  385--390 (1996).

\bibitem{Chun2009}
Chun, M., Wilson, R., Avila, R., Butterley, T., Aviles, J.-L., Wier, D., and Benigni, S., ``{Mauna Kea ground-layer characterization campaign},'' {\em Monthly Notices of the Royal Astronomical Society}~{\bf 394},  1121--1130 (04 2009).

\bibitem{Wildi2003}
{Wildi}, F.~P., {Brusa}, G., {Lloyd-Hart}, M., {Close}, L.~M., and {Riccardi}, A., ``{First light of the 6.5-m MMT adaptive optics system},'' in [{\em Astronomical Adaptive Optics Systems and Applications}{\nolinebreak\hspace{0.1em}]},  {Tyson}, R.~K. and {Lloyd-Hart}, M., eds., {\em Society of Photo-Optical Instrumentation Engineers (SPIE) Conference Series} {\bf 5169},  17--25 (Dec. 2003).

\bibitem{Riccardi2010}
{Riccardi}, A., {Xompero}, M., {Briguglio}, R., {Quir{\'o}s-Pacheco}, F., {Busoni}, L., {Fini}, L., {Puglisi}, A., {Esposito}, S., {Arcidiacono}, C., {Pinna}, E., {Ranfagni}, P., {Salinari}, P., {Brusa}, G., {Demers}, R., {Biasi}, R., and {Gallieni}, D., ``{The adaptive secondary mirror for the Large Binocular Telescope: optical acceptance test and preliminary on-sky commissioning results},'' in [{\em Adaptive Optics Systems II}{\nolinebreak\hspace{0.1em}]},  {Ellerbroek}, B.~L., {Hart}, M., {Hubin}, N., and {Wizinowich}, P.~L., eds., {\em Society of Photo-Optical Instrumentation Engineers (SPIE) Conference Series} {\bf 7736},  77362C (July 2010).

\bibitem{Esposito2010}
{Esposito}, S., {Riccardi}, A., {Fini}, L., {Puglisi}, A.~T., {Pinna}, E., {Xompero}, M., {Briguglio}, R., {Quir{\'o}s-Pacheco}, F., {Stefanini}, P., {Guerra}, J.~C., {Busoni}, L., {Tozzi}, A., {Pieralli}, F., {Agapito}, G., {Brusa-Zappellini}, G., {Demers}, R., {Brynnel}, J., {Arcidiacono}, C., and {Salinari}, P., ``{First light AO (FLAO) system for LBT: final integration, acceptance test in Europe, and preliminary on-sky commissioning results},'' in [{\em Adaptive Optics Systems II}{\nolinebreak\hspace{0.1em}]},  {Ellerbroek}, B.~L., {Hart}, M., {Hubin}, N., and {Wizinowich}, P.~L., eds., {\em Society of Photo-Optical Instrumentation Engineers (SPIE) Conference Series} {\bf 7736},  773609 (July 2010).

\bibitem{Biasi2012}
{Biasi}, R., {Andrighettoni}, M., {Angerer}, G., {Mair}, C., {Pescoller}, D., {Lazzarini}, P., {Anaclerio}, E., {Mantegazza}, M., {Gallieni}, D., {Vernet}, E., {Arsenault}, R., {Madec}, P.~Y., {Duhoux}, P., {Riccardi}, A., {Xompero}, M., {Briguglio}, R., {Manetti}, M., and {Morandini}, M., ``{VLT deformable secondary mirror: integration and electromechanical tests results},'' in [{\em Adaptive Optics Systems III}{\nolinebreak\hspace{0.1em}]},  {Ellerbroek}, B.~L., {Marchetti}, E., and {V{\'e}ran}, J.-P., eds., {\em Society of Photo-Optical Instrumentation Engineers (SPIE) Conference Series} {\bf 8447},  84472G (July 2012).

\bibitem{Close2018}
{Close}, L.~M., {Males}, J.~R., {Morzinski}, K.~M., {Esposito}, S., {Riccardi}, A., {Briguglio}, R., {Follette}, K.~B., {Wu}, Y.-L., {Pinna}, E., {Puglisi}, A., {Xompero}, M., {Quiros}, F., and {Hinz}, P.~M., ``{Status of MagAO and review of astronomical science with visible light adaptive optics},'' in [{\em Adaptive Optics Systems VI}{\nolinebreak\hspace{0.1em}]},  {Close}, L.~M., {Schreiber}, L., and {Schmidt}, D., eds., {\em Society of Photo-Optical Instrumentation Engineers (SPIE) Conference Series} {\bf 10703},  107030L (July 2018).

\bibitem{Christou2014}
{Christou}, J.~C., {Brusa}, G., {Guerra}, J.~C., {Lefebvre}, M., {Miller}, D., {Rahmer}, G., {Sosa}, R., and {Wagner}, M., ``{Living with adaptive secondary mirrors 365/7/24},'' in [{\em Adaptive Optics Systems IV}{\nolinebreak\hspace{0.1em}]},  {Marchetti}, E., {Close}, L.~M., and {Vran}, J.-P., eds., {\em Society of Photo-Optical Instrumentation Engineers (SPIE) Conference Series} {\bf 9148},  91480F (Aug. 2014).

\bibitem{Chun2022}
{Chun}, M.~R., {Ryan}, A., {Zhang}, R., {Kuiper}, S., {Ackaert}, G., {Baranec}, C., {Baeton}, M., {Bos}, A., {Bowens-Rubin}, R., {Dekker}, B., {Dungee}, R., {Gupta}, T., {Hinz}, P., {Jonker}, W., {Kamphues}, F., {Lai}, O., {Lu}, J., {Maniscalco}, M., {Monna}, B., {Nair}, M., {Nigenhuis}, J., {Priem}, H., and {Vogel}, P.-A., ``{Progress on the University of Hawaii 2.2-meter adaptive secondary mirror},'' {\em Society of Photo-Optical Instrumentation Engineers (SPIE) Conference Series} (2022).

\bibitem{Jonker2022}
{Jonker}, W.~A., {Vogel}, P.-A., {Horst}, R.~t., {Vu}, A.~T., {Kuiper}, S., and {Kamphues}, F., ``{Hot forming of a large deformable mirror facesheet},'' {\em Society of Photo-Optical Instrumentation Engineers (SPIE) Conference Series} (2022).

\bibitem{BRubin2021}
{Bowens-Rubin}, R., {Hinz}, P., {Kuiper}, S., and {Dillon}, D., ``{Performance of large-format deformable mirrors constructed with hybrid variable reluctance actuators II: initial lab results from FLASH},'' in [{\em Techniques and Instrumentation for Detection of Exoplanets X}{\nolinebreak\hspace{0.1em}]},  {Shaklan}, S.~B. and {Ruane}, G.~J., eds., {\em Society of Photo-Optical Instrumentation Engineers (SPIE) Conference Series} {\bf 11823},  118231R (Sept. 2021).

\bibitem{Lee2024}
{Lee}, E., {Chun}, M.~R., {Lai}, O., {Zhang}, R., {Baeten}, M., {Bos}, A., {Kidron}, M., {Kamphues}, F., {Kuiper}, S., {Jonker}, W., {Connelley}, M.~S., {Rayner}, J., {Ryan}, A., {Hinz}, P.~M., {Bowens-Rubin}, R., {Lockhart}, C., and {Kelii}, M., ``{First laboratory and on-sky results of an adaptive secondary mirror with TNO-style actuators on the NASA Infrared Telescope Facility},'' {\em SPIE}, International Society for Optics and Photonics (2024).

\bibitem{Bos2024}
{Bos}, A., {Kuiper}, S., {Kidron}, M., {Dekker}, B., {Baeten}, M., {Salcedo}, E.~V., {Vermeulen}, R., {Kuijt}, J., {Boot}, K., {Venrooy}, B.~v., {Buuren}, R.~v., {Kamphues}, F., {Jonker}, W.~A., {Maniscalco}, M., {Chun}, M.~R., {Connelley}, M.~S., {Vleggaar}, J.~J., and {Hinz}, P.~M., ``{Status and first results of the NASA IRTF Adaptive Secondary Mirror},'' {\em SPIE}, International Society for Optics and Photonics (2024).

\bibitem{Noll1976}
{Noll}, R.~J., ``{Zernike polynomials and atmospheric turbulence.},'' {\em Journal of the Optical Society of America (1917-1983)}~{\bf 66},  207--211 (Mar. 1976).

\bibitem{kolmogorov}
{Kolmogorov}, A.~N., ``{The Local Structure of Turbulence in Incompressible Viscous Fluid for Very Large Reynolds Numbers},'' {\em Proceedings of the Royal Society of London Series A}~{\bf 434},  9--13 (July 1991).

\bibitem{Baranec2024}
International Society for Optics and Photonics,  [{\em {Commissioning results from the Robo-AO-2 facility for rapid visible/near-infrared AO imaging}}{\nolinebreak\hspace{0.1em}]}, SPIE (2024).

\end{thebibliography}
\bibliographystyle{spiebib} 

\end{document}